\documentclass[fleqn,usenatbib]{mnras}

\usepackage{hyperref}
\usepackage{siunitx}
\usepackage{amsmath}
\usepackage{booktabs}
\let\oldAA\AA
\renewcommand{\AA}{\text{\normalfont\oldAA}}

\usepackage[T1]{fontenc}
\usepackage{xcolor}
\DeclareRobustCommand{\VAN}[3]{#2}
\let\VANthebibliography\thebibliography
\def\thebibliography{\DeclareRobustCommand{\VAN}[3]{##3}\VANthebibliography}

\usepackage{graphicx}
\usepackage{grffile}
\usepackage{amssymb}
\usepackage{newtxtext,newtxmath}

\renewcommand{\vec}[1]{ {\bf #1}}
\usepackage{xcolor}

\title[Cosmology behind the mask]{Cosmology behind the mask: Constraining the parameters of $\Lambda$CDM with the unmasked galaxy density field from VIPERS}

\author[N. Estrada et al.]{
N. Estrada,$^{1,2}$\thanks{E-mail: gilbertonicolas.estradamartinez@phd.unipd.it}
B.R. Granett,$^{3}$
L. Guzzo$^{4,3,5}$
\\
$^{1}$Dipartimento di Fisica e Astronomia ``Galileo Galilei'', Universit\`{a} degli Studi di Padova, Vicolo dell'Osservatorio, 3, I-35122, Padova, Italy\\
$^{2}$INAF - Osservatorio Astronomico di Padova, Vicolo dell'Osservatorio, 5, I-35122, Padova, Italy \\
$^{3}$INAF - Osservatorio Astronomico di Brera, Via Brera 28, 20122 Milano – via E. Bianchi 46, 23807 Merate, Italy \\
$^{4}$Dipartmento di Fisica, Universit\`{a} degli Studi di Milano, Via G. Celoria, 16, I-20133, Milano, Italy \\
$^{5}$INFN - Sezione di Milano, via G. Celoria 16, 20133 Milano, Italy
}

\date{Accepted XXX. Received YYY; in original form ZZZ}

\pubyear{2021}

\begin{document}
\label{firstpage}
\pagerange{\pageref{firstpage}--\pageref{lastpage}}

\maketitle

\begin{abstract}
Galaxy redshift surveys are designed to map cosmic structures in three dimensions for large-scale structure studies. Nevertheless, limitations due to sampling and the survey window are unavoidable and degrade the cosmological constraints.
We present an analysis of the VIMOS Public Extragalactic Redshift Survey (VIPERS) over the redshift range $0.6 < z < 1$ that is optimised to extract the cosmological parameters while fully accounting for the complex survey geometry. We employ the Gibbs sampling algorithm to iteratively draw samples of the galaxy density field in redshift space, the galaxy bias, the matter density, baryon fraction and growth-rate parameter $f\sigma_8$ based on a multivariate Gaussian likelihood and prior on the density field. Despite the high number of degrees of freedom, the samples converge to the joint posterior distribution and give self-consistent constraints on the model parameters. We validate the approach using VIPERS mock galaxy catalogues. Although the uncertainty is underestimated by the Gaussian likelihood on the scales that we consider by 50\%, the dispersion of the results from the mock catalogues gives a robust error estimate. We find that the precision of the results matches those of the traditional analyses applied to the VIPERS data that use more constrained models. By relaxing the model assumptions, we confirm that the data deliver consistent constraints on the $\Lambda$CDM model. This work provides a case-study for the application of maximum-likelihood analyses for the next generation of galaxy redshift surveys.

\end{abstract}



\section{Introduction}

According to the standard model of cosmology, the cosmic web of galaxies consisting of groups, clusters and filaments that we observe today grew from nearly homogeneous conditions in the early Universe.  The formation of structure through gravitational collapse over cosmic history is governed by two puzzling elements in the standard model: dark matter and dark energy. While dark matter drives the formation of structure, dark energy is responsible for the acceleration of the expansion of the Universe at late times and slows the rate of structure formation. The search for a physical interpretation of these components has motivated the latest generation of galaxy surveys \citep{euclid,desi,lsst,des}.

Despite the formation of complex structures, the distribution of matter on large scales was established at the epoch of inflation in the early Universe and is well described by a Gaussian field. The field is specified by its first and second moments: the mean density and the power spectrum of the Fourier mode amplitudes: $P(k)\equiv \langle \delta_k^2 \rangle$. Measurements of the shape of the power spectrum allow us to constrain the abundances of the components of the Universe, namely baryonic matter, dark
matter and neutrinos \citep{Alam17,rota17,cole05,tegmark04}.  Moreover, the baryon acoustic oscillations in the power spectrum provide a standard ruler to measure the expansion history and constrain the properties of dark energy \citep{eisenstein06}. 

Galaxy redshift surveys are sensitive to the velocity field projected along the line of sight due to the effect of redshift-space distortions \citep{kaiser87}. The anisotropic signal encodes information about the logarithmic growth rate of structure and provides a key test of the gravity model on cosmological scales \citep{Guzzo08,Percival2009}. 

The power spectrum estimate for a Gaussian random field has a formal error due to sample variance of $\sigma_P^2 = 2/N_{\rm modes} P(k)$ \citep{tegmark97}. However, even in the ideal case of a Gaussian field this limit cannot be achieved due to the loss of information inherent in the survey selection process, namely the sampling process and the window due to the finite survey volume. For a Poisson noise model, the FKP \citep{feldman94} estimator minimizes the loss of information due to discrete sampling. The estimator can be further optimized by incorporating galaxy weights that account for the relationship between the amplitude of the power spectrum and galaxy properties \citep{MonteroDorta20,Pearson16,Cai11,Percival04}.

We can aim to extract the maximal information present in a survey by maximising the likelihood of the multi-variate Gaussian density field measured on a grid \citep{Tegmark97b}:
\begin{equation}
    L \propto \det|C|^{-1/2} \exp\left(-\frac{1}{2} \sum_{ij} \delta_i C^{-1}_{ij} \delta_j \right).
\end{equation}
This expression gives the likelihood of measuring pairs of grid cells with over-densities $\delta_i$ and $\delta_j$ related by the covariance $C_{ij}$. The covariance matrix $C$ fully characterises the information in the underlying Gaussian field and is related to the two-point correlation function in configuration space or the power spectrum in Fourier space. The covariance depends on the cosmological model through a set of parameters $C=C(\{\theta\})$ that we aim to estimate. The error on the maximum-likelihood solution  is related to the second derivative of the likelihood with respect to the model parameters:
\begin{equation}
\sigma_{\textcolor{black}{\alpha\beta}}^2 = \left\langle \frac{d^2 \log L}{d \theta_{\textcolor{black}{\alpha}} d \theta_{\textcolor{black}{\beta}}} \right\rangle^{-1}. 
\end{equation}

This expression gives the covariance between the estimates of two parameters $\theta_\alpha$ and $\theta_\beta$. The angle brackets indicate the ensemble average over realizations of the field.
The maximum likelihood solution promises to give the minimum variance estimate of the parameters in the presence of a complex survey geometry and galaxy selection function. In practice, since the galaxy density field is not Gaussian, the formal error may not be achieved; nevertheless, we expect that the optimisation accounting for the selection function and survey geometry to hold and provide important gains over the standard analysis techniques.

The application of the maximum-likelihood formalism to galaxy surveys has been limited due to the computational requirements. \citet{Philcox2020} present a modern implementation of the quadratic estimator. In the past, quadratic estimators have been applied in angular galaxy clustering analyses \citep{Granett12,Ho12,2001MNRAS.325.1603E}. Applications to the three-dimensional density fields include the Karhunen-Loeve decomposition implemented on the Sloan Digital Sky Survey \citep{Pope04}. \textcolor{black}{Maximum-likelihood estimators applied to large-scale structure were subsequently generalised with the Bayesian estimation framework \citep{Lavaux16, Jasche10, Kitaura08}.}

In this work we develop a novel estimator that solves the maximum-likelihood problem by Monte Carlo sampling. We test the approach using the VIMOS Public Extragalactic Redshift Survey \citep[VIPERS,][]{scodeggio16,guzzo14} and simulated galaxy catalogues.  VIPERS measured the redshifts of approximately $10^5$ galaxies at a median redshift $z \simeq 0.7$ and produced an unprecedented dataset for cosmological studies at intermediate redshifts. The combination of broad galaxy selection function, volume and high sampling rate is unique and enables novel cosmological analyses.

In \citet{granett15}, we carried out the first maximum-likelihood analysis of the VIPERS density field on a subset of the final data release. That analysis focused on optimising the measurement of the matter power spectrum accounting for the dependence of the galaxy bias on galaxy luminosity and colour. We extend that work here using the final VIPERS data release (PDR2) and carry out a \textcolor{black}{general} analysis of cosmological parameters. \textcolor{black}{In this work we marginalise over the density field to sample as free parameters: matter density (sampled as $\Omega_Mh$, with $h$ fixed), baryon fraction ($f_b = \Omega_B/\Omega_M$), galaxy bias ($b$), velocity dispersion ($\sigma_v$) and the combination of $\sigma_8$ and the growth rate ($f\sigma_8$). Priors on those parameters are given in table \ref{table:prior}.} 

The first complete cosmological analysis of VIPERS was presented in \cite{rota17}. This work analysed the galaxy power spectrum monopole using the FKP estimator and constrained the matter density and baryon fraction. The baryon acoustic oscillation signal was not detected in the power spectrum measurement due to the effect of the survey window function.  Here, we make a direct comparison using mock galaxy catalogues between the constraints \textcolor{black}{on $\Omega_m$ and $\Omega_b$} from \cite{rota17} and the maximum likelihood technique. 

\textcolor{black}{The growth rate in combination with the linear clustering amplitude, $f\sigma_8$, was measured from VIPERS based on the analysis of the correlation function multipoles \citep{pezzotta17,mohammad18} and the void-galaxy correlation function \citep{Hawken17}. Inferring the growth rate $f\equiv -d\log \sigma_8(z)/d\log  (1+z)$ alone requires breaking the degeneracy between $\sigma_8$ and galaxy bias and was carried out with VIPERS through the joint analysis of the correlation function multipoles with galaxy-galaxy lensing \citep{Delatorre17} and three-point correlation statistics \citep{Veropalumbo2021}.}

\textcolor{black}{Here we limit the analysis to two-point clustering, so we are restricted to analysing the parameter combination $f\sigma_8$. }Compared with \textcolor{black}{the} previous \textcolor{black}{VIPERS} studies \textcolor{black}{based on the galaxy power spectrum and multipoles of the correlation function}, the analysis we present here has a greater freedom in the model since we simultaneously fit the matter density parameters and \textcolor{black}{f$\sigma_8$}. The results will be contrasted with the previous VIPERS analyses in Sec.~\ref{sec:conclusions}.

This paper begins in Section~\ref{sec:VIPERS} with a brief description of the VIPERS survey and its mask. In Section~\ref{sec:datamodel} we define the physical quantities that we aim to obtain and introduce the Gibbs sampling formalism. The algorithm developed for this work is executed on VIPERS data and tested on mock catalogues and the results are presented in Section~\ref{sec:results}. A summary of the results and a discussion of their validity is found in Section~\ref{sec:conclusions}.

We assume \textcolor{black}{a $\Lambda$CDM flat model using the \cite{planck18} with} $A_s=2.0968 \times 10^{-9}$, $n_s=0.9652$, $Y_{He}=0.2454$, $H_0=0.6732$.  \textcolor{black}{$\Omega_M$ and $\Omega_b$ are free parameters}. Magnitudes are in the AB system unless noted. 

\section{Experimental Data: VIPERS}
\label{sec:VIPERS}
\begin{figure}
	\includegraphics[width=\columnwidth]{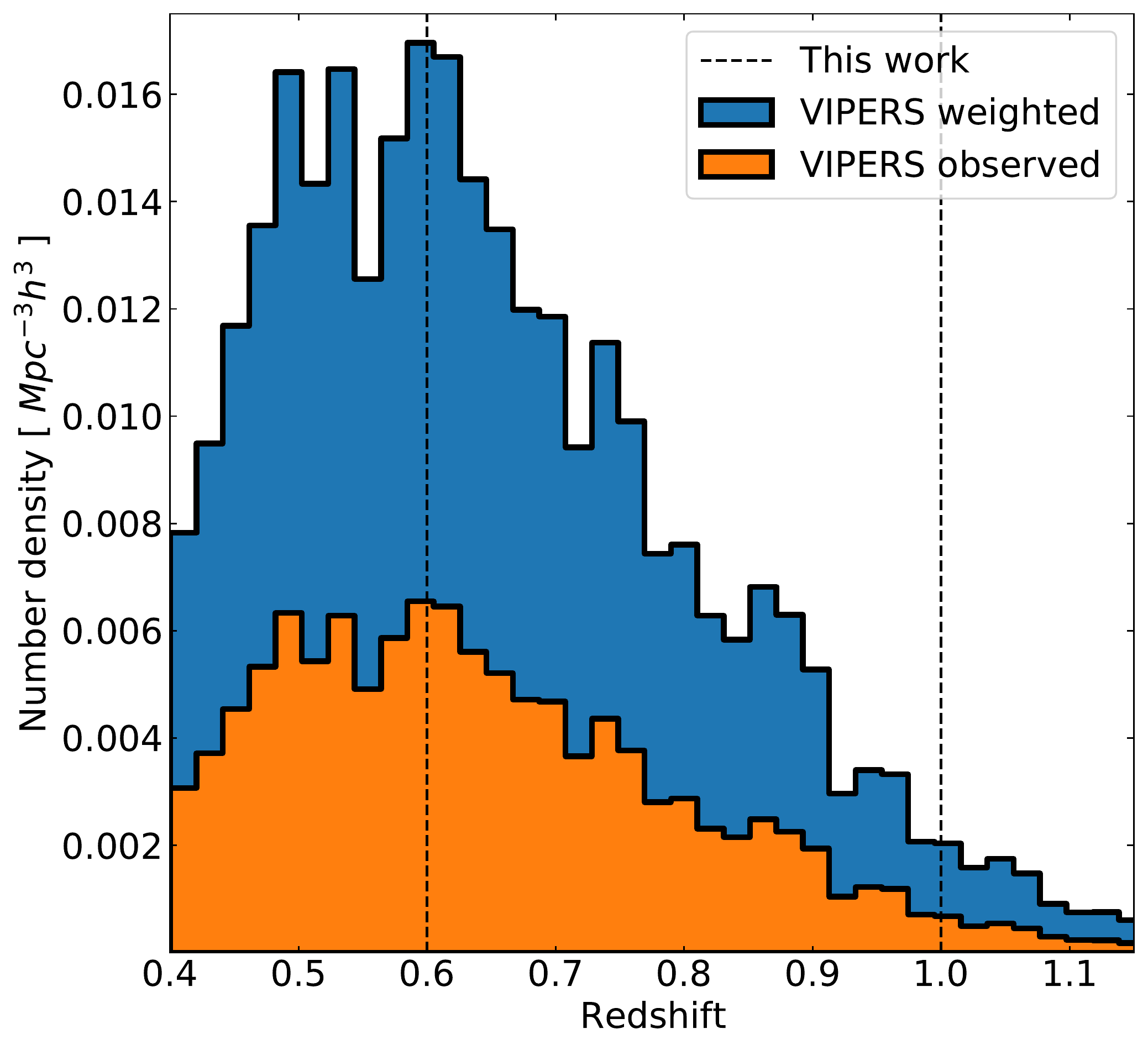}
    \caption{Mean spatial density of galaxies as a function of redshift for the final VIPERS sampled used in this work. The lower histogram is the observed distribution, while the top one show the effect of the completeness corrections including the spectroscopic success rate (SSR) and target sampling rate (TSR). The vertical dashed lines define the boundaries of the redshift range 0.6-1.0 used in this analysis.}
    \label{fig:redshift}
\end{figure}

\subsection{Sample Selection}
\label{sec:sampleselection}
The VIMOS Public Extragalactic Redshift Survey (VIPERS) is an ESO programme on VLT \citep[European Southern Observatory - Very Large Telescope;][]{guzzo14,scodeggio16}. The survey targeted galaxies for medium resolution spectroscopy using VIMOS \citep[VIsible Multi-Object Spectrograph;][]{lefevre03} within two regions of the W1 and W4 fields of the CFHTLS-Wide Survey \citep[Canada-France-Hawaii Telescope Legacy Wide;][]{cuillandre12}. Targets were chosen with a limiting flux with $i_{AB} < 22.5$ based upon a colour selection to be in the redshift range 0.5 - 1.2. The overall footstep of VIPERS is 23.5 $deg^2$, corresponding to an effective sky coverage of 16.3 $deg^2$, with 97\,414 spectroscopically observed galaxies giving a peak number density of $\Bar{n} \sim 6 \times 10^{-3} h^3Mpc^{-3}$ at $z\sim0.6$.

We used the second and final VIPERS public data release  \citep[PDR2,][]{scodeggio16}. The redshift distribution is shown in Fig. \ref{fig:redshift}. As in previous statistical analyses of the VIPERS dataset, we used only galaxies with secure redshift measurements, defined as having a quality flag between 2 and 9 inclusive, corresponding to an overall redshift confirmation rate of 98\% \citep[see][for definitions]{guzzo14}.

We selected sources from the VIPERS catalogue in the redshift range $0.6 < z <  1.0$. The lower bound at $z = 0.6$ fully excludes the transition region produced by the nominal $z = 0.5$ colour-colour cut of VIPERS. The high redshift limit at $z=1.0$ excludes the most sparse distant part of the survey, where shot noise dominates and so the effective volume is small.  The total number of sources used in this work is 73\,572.

\subsection{Survey Mask}
\label{sec:mask}

The VIPERS survey selection function is characterised by an angular mask and completeness weights for each galaxy \citep{guzzo14,scodeggio16}. The completeness is the product of the target sampling rate (TSR) and the spectroscopic measurement success rate (SSR): $c=c_{TSR} c_{SSR}$.

The target sampling rate (TSR) accounts for the selection of targets for spectroscopy. The slit assignment for the VIMOS spectrograph was optimised using the SSPOC algorithm with the primary constraint that spectra cannot overlap on the focal plane \citep{bottini2005}. This led to a suppression in the clustering amplitude on large scales. This suppression is corrected for on large scales by the factor  $1/c_{TSR}$ where $c_{TSR}$ depends on the \textcolor{black}{local} density of sources on the sky \citep{pezzotta17}.

The second factor in the completeness accounts for the spectroscopic measurement success rate (SSR).  The success rate depends primarily on the observational conditions and the flux of \textcolor{black}{each}  source \citep{scodeggio16}. \textcolor{black}{The TSR and SSR are computed for each galaxy in the survey.} The combined galaxy weight $w = (c_{TSR} c_{SSR})^{-1}$ corrects for both the TSR and SSR effects. The weighted redshift distribution is shown in Fig. \ref{fig:redshift}.  \textcolor{black}{We define the effective redshift of the sample using the distribution of galaxy pairs which can be approximated based on the weighted redshift distribution, $n(z)$, in the range $0.6<z<1.0$, as,}
\begin{equation}
    z_{\rm eff} = \frac{\sum_i n_i^2 z_i}{\sum_i n_i^2}\,.
\end{equation}
\textcolor{black}{with bin size $\Delta z=0.02$ and where $i$ indexes the redshift bins and $z_i$ is the midpoint of the bin. The resulting effective redshift of the sample is $z_{\rm eff}=0.71$.}

\subsection{Mock Catalogues}
\label{sec:mockcatalogues}
We used a set of simulated VIPERS galaxy catalogues to test the algorithm developed for this work and quantify the level of systematic biases in the final results. These mock catalogues were built using the MultiDark N-body simulation \citep[BigMD,][]{klypin16} with the Planck cosmology ($\Omega_M$, $\Omega_{\Lambda}$, $\Omega_B$, $h$, $n_s$, $\sigma_8$) = (0.307, 0.693, 0.0482, 0.678, 0.960, 0.823). A halo occupation distribution (HOD) prescription was applied to add galaxies to the dark matter haloes which was calibrated from the VIPERS data \citep{Delatorre13}. We used 153 independent mock catalogues in this work.

The mock catalogues match the number density of the VIPERS catalogue after correcting for the TSR and SSR selection effects. They reproduce the angular survey mask but do not include the spectrograph slit placement constraints. Therefore the TSR and SSR completeness weights are not applied in the mock analyses. The redshift measurement error was drawn from a Gaussian distribution with standard deviation $\sigma_z=0.00054(1+z)$.

\section{Data Model}
\label{sec:datamodel}

\subsection{Density field}
\label{sec:densityfield}
We consider the galaxy sample projected onto a comoving cartesian grid \textcolor{black}{with cubic cells of 5 $Mpc \, h^{-1}$ side. The dimensions of the grid are $77\times16\times184$ cells, roughly corresponding to RA $\times$ DEC $\times$ redshift. This size was determined to contain the largest possible light cone corresponding to the VIPERS W1 field within the comoving distances allowed by the prior. The grid is calibrated to contain the W1 field, which is the larger one, but an identical grid is used for the W4 field, to consistently compute the Fourier Transfom with the same grid dimensions}. The number of galaxies observed in a given grid cell is parametrised by
\begin{equation}
    n_{obs,i} = \bar{n}_i (1 + \delta_{g,i}) + \epsilon_i,
\end{equation}
where $\bar{n}_i$ is the expected number in the cell given the survey selection function and $\delta_{g,i}$ is the underlying galaxy over-density field in redshift space and $\epsilon_i$ is the noise contribution. \textcolor{black}{$n_{obs,i}$ is computed inside the survey volume according to the nearest gridpoint assignment scheme, with zeros outside the survey area. We assume that $n_{obs,i}$ is a Poisson process such that the variance is $\sigma^2_{\epsilon,i} = \langle \epsilon_i^2 \rangle = \bar{n}_i$. The mean number of objects in the cell, $\bar{n}_i$, was generated using a random catalogue which was built following the VIPERS cumulative redshift distribution. The redshift distribution weighted by the inverse of the completeness was combined from the two fields W1 and W4 and smoothed with a Gaussian kernel with width in redshift of $\sigma=0.07$.}
The matter density field is related to the galaxy field by a constant bias factor $b$: $\delta_g = b\delta$ \textcolor{black}{which is evaluated at the effective redshift. The use of an effective redshift and bias to model the power spectrum (see Eq. \ref{eq:modelpowerspectrum}) is warranted because the amplitude parameterised by $b(z)\sigma_8(z)$ does not evolve significantly in the VIPERS sample, as was shown in \citet{rota17}.}

The statistical properties of the density field $\delta$ are described by the power spectrum of the Fourier modes: $P_i = \langle | \tilde \delta_i |^2 \rangle$. The Fourier transform of $\delta$ is defined by
\begin{equation}
\tilde \delta(\vec{k}) = \frac{1}{2\pi}\int d^3\vec{x} \delta(\vec x) e^{-i \vec{k} \vec{x}},
\end{equation}
which we evaluate on a discrete coordinate grid using the fast Fourier transform (FFT) algorithm.

In the \textcolor{black}{flat }$\Lambda$CDM model, the \textcolor{black}{linear} matter power spectrum $P_m(k)$ depends on redshift, $\Omega_m$, $\Omega_b$, $H_0$ \textcolor{black}{and $\sigma_8$}. We adopt the non-linear model of the matter power spectrum using the CLASS code with the Halofit prescription \citep{Blas2011,Takahashi2012}.

In order to model the observations we parametrise the anisotropic redshift-space power spectrum using the dispersion model, i.e. the Kaiser linear expression modified to empirically account for the velocity dispersion \citep{peacock94}:
\begin{equation}
    P^s(\vec k) = \left( b + f \mu^2 \right)^2 e^{-k_{||}^2 \sigma_{v}^2} P_{m}(|\vec k|)\,.
    \label{eq:modelpowerspectrum}
\end{equation}
The free parameters are the galaxy bias $b$, the logarithmic growth rate $f$ and the velocity dispersion $\sigma_{v}$, while the nonlinear matter power spectrum $P_{m}(k)$ depends on the cosmological parameters. The effective velocity dispersion term $\sigma_{v}$ accounts for the sum of both the true pairwise dispersion along the line of sight and the redshift measurement error, which sum in quadrature. The parameter $\mu = k_{||}/k$ is the cosine of the angle of the wavevector $\vec{k}$ with respect to the line-of-sight.

Since we estimate the density field on a discrete grid, the power spectrum is affected by aliasing \citep{jing05}. We include the effect arising from the nearest-grid point mass assignment in the power spectrum model:
\begin{equation}
P_{grid}(\vec k) = \sum_n |W(\vec{k} + 2k_N\vec{n})|^2 P^s(\vec{k} + 2k_N\vec{n}),
\label{eq:powergrid}
\end{equation}
where the summation includes the first harmonic so the three-dimension grid $\vec n = (n_x, n_y, n_z)$ where each element takes integer values -1, 0 and 1. The window function $W$ corresponds to the \textit{sinc} function for the nearest grid point mass assignment scheme  
\textcolor{black}{This treatment differs from \citet{granett15} that did not include an aliasing correction but reduced the effect by using super-sampling when binning galaxies onto the grid. Here we save computation time when building the grid, and instead evaluate the aliasing terms in the likelihood.} A shot noise term is not included in the power spectrum model, but instead a shot noise correction is included in the estimator (see Sec \ref{sec:wienerfilter}).

\subsection{Estimator}
We construct an estimator for the model parameters starting from the observed galaxy number counts. From Bayes' theorem we write the posterior distribution function of the set of cosmological parameters $\vec{\Omega}$ as
\begin{equation}\label{eq:one}
    p(\vec{\Omega} |  \vec{n}_{obs}) \propto p(\vec{n}_{obs} | \vec{\Omega})\ p(\vec{\Omega}) .
\end{equation}
The expectation of the posterior distribution provides an estimate of the parameters:
\begin{equation}
    \langle \vec{\Omega} \rangle = \int \vec{\Omega} p(\vec{\Omega} | \vec{n}_{obs}) d\vec{\Omega}.
\end{equation}
To evaluate this, we introduce the underlying density field as a latent variable.  Marginalizing over this field $\vec{\delta}$, we write,
\begin{eqnarray}
    p(\vec{\Omega} |  \vec{n}_{obs}) &=& \int p(\vec{\Omega} |  \vec{n}_{obs}, \vec{\delta}) p(\vec{\delta}) d\vec{\delta} \\
    &=& \int p(\vec{\Omega}, \vec{\delta}|  \vec{n}_{obs}) d\vec{\delta}.
\end{eqnarray}
The integrand can be written as
\begin{equation}\label{eq:three}
    p(\vec{\Omega},\vec{\delta} |  \vec{n}_{obs} ) = p(\vec{n}_{obs} | \vec{\delta})\ p(\vec{\delta} | \vec{\Omega}) p(\vec{\Omega}).
\end{equation}
The first term $p(\vec{n}_{obs} | \vec{\delta})$ is the data likelihood, which we will describe by a multi-variate Gaussian distribution:
\begin{equation}
  p(\vec{n}_{obs} | \vec{\delta}) = \left((2\pi)^N \Pi_i \bar{n}_i \right)^{-\frac{1}{2}} \exp\left(-\frac{1}{2}\sum_i \frac{\left[n_{obs,i} - \bar{n}_i (1 + b\delta)\right]^2}{\bar{n}_i}\right).
\end{equation}
The second term $p(\vec{\delta} | \vec{\Omega})$ represents the Bayesian prior. We consider the power spectrum of $\delta$ evaluated on the grid, $P_{grid,i} = \langle |\tilde\delta_i|^2\rangle$ which depends implicitly on the model parameters: $P_{grid,i} \equiv P_{grid,i}( \vec{\Omega})$. This formulation results in a Gaussian prior on $\delta$:
\begin{equation}
p(\vec{\delta} | \vec{\Omega}) = \left((2\pi)^N \Pi_i P_{grid,i} \right)^{-\frac{1}{2}}\exp\left(-\frac{1}{2}\sum_{i\textcolor{black}{<k_{max}}} \frac{|\tilde{\delta}_{i}|^2}{P_{grid,i}} \right).
\label{eq:likelihoodemcee}
\end{equation}

\textcolor{black}{The Fourier transform of the density field $\tilde\delta$ was computed with by FFT without zero padding.  In practice we did not evaluate the prior using all modes to the Nyquist frequency but  applied a limit, $k_{max}$.  In the analysis we set $k_{max}=0.4 h Mpc^{-1}$ and the Nyquist frequency is $k_N=0.6 h Mpc^{-1}$.}

\subsection{Gibbs sampler}
The high dimensionality of the posterior distribution 
\begin{equation}
p(\vec{\delta}, b, f\textcolor{black}{\sigma_8}, \sigma_v, \Omega_m\textcolor{black}{h}, \textcolor{black}{f_b} | \vec{n}_{obs}),
\label{eq:prior}\end{equation}
makes it challenging to sample from directly. We employ the Gibbs sampling algorithm to divide the problem into two steps. We first draw a realization of the density field $\delta$ with the parameters fixed \textcolor{black}{through the Wiener filter (\ref{eq:gibbs1})}, and next sample the parameters with $\delta$ fixed \textcolor{black}{through a secondary MCMC chain (\ref{eq:gibbs2})}. Iterating these two steps in sequence allows us to draw samples from the joint distribution.
Schematically, the two steps are written below.
\begin{equation}
    \tag{i}
    \delta_{\gamma+1} \leftarrow p(\delta | \vec n_{obs}, P_\gamma) \label{eq:gibbs1},
\end{equation}

\begin{equation}
    \tag{ii}
    \left(\begin{array}{c}b_{\gamma+1}\\ f\sigma_{8\gamma+1}\\ \sigma_{v \gamma+1}\\ \Omega_mh_{\gamma+1}\\  f_{b\gamma+1}\end{array}\right) \leftarrow p(b, f\sigma_8, \sigma_v, \Omega_mh, f_b | \delta_{\gamma+1}) \label{eq:gibbs2},
\end{equation}
\textcolor{black}{where $P_{\gamma}$ is the power spectrum computed at the step $\gamma$.This approach differs from the sampling scheme implemented in \citep{granett15}. In that work we sampled bins of the power spectrum and only fit the cosmological parameters after sampling. Instead, here, the power spectrum is parametrised by Eq. \ref{eq:modelpowerspectrum} and evaluated with the CLASS code on every iteration. This approach avoids binning the power spectrum, since it is evaluated on the Fourier grid, and guarantees self-consistency of the power spectrum estimation and cosmological parameter constraints.}

\subsection{Wiener Filter}
\label{sec:wienerfilter}
The first Gibbs sampling step \ref{eq:gibbs1} corresponds to generating a density field that maximizes the posterior distribution $p(\delta | \vec n_{obs}, P_\gamma)$, which is given by the Wiener filter  \citep{rybickipress92}. The solution $\delta^{WF}$ is found by solving the linear equation:
\begin{equation}
    \sum_j \big(S_{ij}^{-1}+N_{ij}^{-1}\big) \, \, \delta^{WF}_j =  \sum_j N_{ij}^{-1} \, \, \delta_{g,j},
    \label{eq:wiener}
\end{equation}
here, $S$ and $N$ are the signal and noise covariance matrices. Under the model assumptions, the signal is diagonal in Fourier space and can be written in terms of the power spectrum: $\tilde{S}_{ii}=P_i$. Conversely, the noise matrix is diagonal in configuration space: $N_{ii}=\bar{n}^{-1}$. As $P_i$ and $\bar{n}_i$ are diagonal matrices in different bases, it is computationally expensive to operate simultaneously with them and compute the inverse of their sum. Instead, we apply the iterative conjugate gradient solver to estimate the solution $\delta^{WF}_i$. In practice, we first compute the product $P_i^{-1} \, \delta^{WF}_i$ in Fourier space, and by using an inverse Fourier transform, return to configuration space, where $\bar{n}_i$ and $\delta_{g,i}$ are diagonal matrices.

While the solution $\delta^{WF}$ maximizes the posterior function and therefore carries the full information of the density field, it lacks power on small scales. In order to sample a full realization  from the posterior distribution, we construct the constrained fluctuation field  $\chi$ which is uncorrelated with $\delta^{WF}$. The constrained realization  $\delta = \delta^{WF} + \chi$ samples from the posterior distribution and recovers the target power spectrum. The constrained fluctuation field is found by solving the equation \citep{jewell04}:
\begin{equation}
    \sum_j \big(S_{ij}^{-1}+N_{ij}^{-1}\big) \, \, \chi_j =  \sum_j\big( S_{ij}^{-1/2} \xi_{1,j} + N_{ij}^{-1/2} \xi_{2,j}\big),
    \label{eq:xiwiener}
\end{equation}
where $\xi_1$ and $\xi_2$ are uniform random fields between 0 and 1 in configuration space. We solve Eq \ref{eq:xiwiener} using the conjugate gradient method.

\subsection{Likelihood on Cosmological Parameters}
\label{sec:emcee}
The second Gibbs sampling step \ref{eq:gibbs2} corresponds to obtaining a new set of parameters $\Omega$ based on a given realization of the density field $\delta$. The posterior distribution of the parameters is 
\begin{equation}
    p(\Omega | \delta) = p(\delta | \Omega) p(\Omega),
    \label{eq:secondbayesgibbs}
\end{equation}
where the conditional probability $p(\delta | \Omega) $ is computed with Eq. \ref{eq:likelihoodemcee}. 

To integrate over the posterior distribution $p(\Omega|\delta)$, we employed a Markov Chain Monte Carlo (MCMC) algorithm based on the {\tt Emcee} ensemble sampler \citep{foreman2013_emcee}. The likelihood $p(\delta|\Omega)$ provided to the {\tt Emcee} code is described in equation \ref{eq:likelihoodemcee}. We use a flat prior $p(\Omega)$ on the parameters, described in table \ref{table:prior}, which allows the walkers to explore completely the physical configurations of the system. \textcolor{black}{In this work, different from \citet{granett15}, all parameters are sampled jointly.}

To explore the five-dimensional parameter space, the {\tt Emcee} ensemble sampler was configured with 20 walkers and 20 steps for each walker, starting from randomly selected points from within the prior. The final position of one of these walkers was employed as the input for next step of the Gibbs chain, allowing the algorithm to explore the whole prior region. 

\begin{table}
\centering
\begin{tabular}{c c}
\hline\hline
\textit{Parameter} & \textit{Prior} \\
\hline
Matter density & 0.1 $ < \Omega_Mh < $ 0.3  \\
Baryon fraction & 0  $< f_B <$ 0.3 \\
Galaxy bias & 1  $< b <$  2 \\
Velocity dispersion & 1.5  $< \sigma_v <$  3.5 \\
Linear growth rate & 0.25  $< f\sigma_8 <$  0.55 \\
\hline
\end{tabular}
\caption{\label{table:prior}Flat prior on \textcolor{black}{the} cosmological parameters \textcolor{black}{sampled in this work}. The walkers described in section \ref{sec:emcee} are free to move within these intervals.}
\end{table}

\section{Application to data}
\label{sec:results}
\begin{figure*}
\includegraphics[width=18cm]{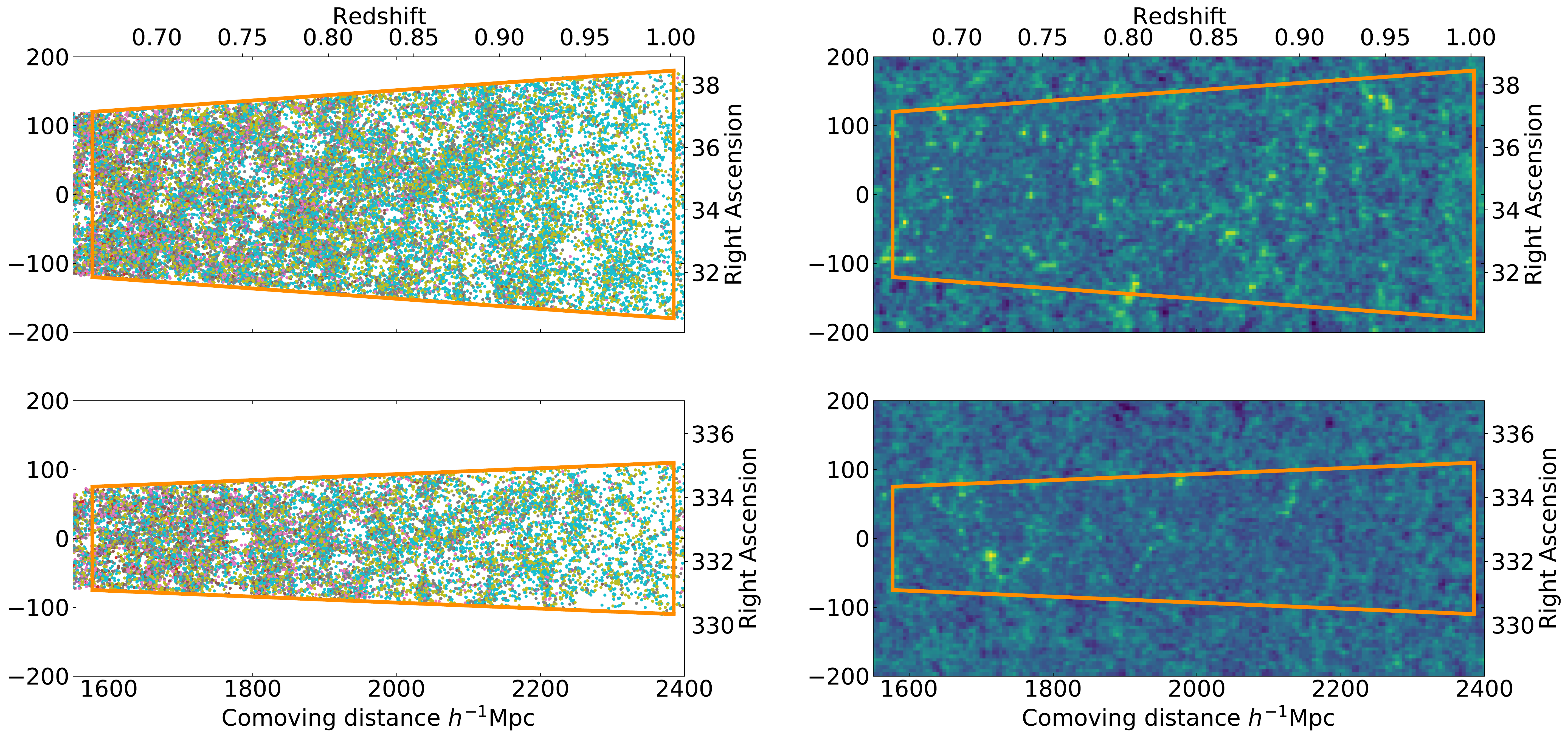}
\caption{VIPERS cone diagrams for the fields W1 (top) and W4 (bottom). \textit{Left}: redshift-space positions of observed galaxies. The orange line traces the field boundaries cut in the redshift direction at 0.6 < z < 1.0. The graph includes all galaxies projected along the declination. Each galaxy is represented by a filled circle coloured according to its \textit{i} band luminosity. \textit{Right}: a 10 $h^{-1} Mpc$  slice of the reconstructed density field taken from one step of the Gibbs sampler. It represents the anisotropic Wiener reconstruction from the weighted combination of galaxy tracers. The field is filled with a constrained Gaussian realisation in the volume obscured by the survey mask.}
\label{fig:viperscone}
\end{figure*}
\begin{figure*}
\center
	\includegraphics[width=14cm]{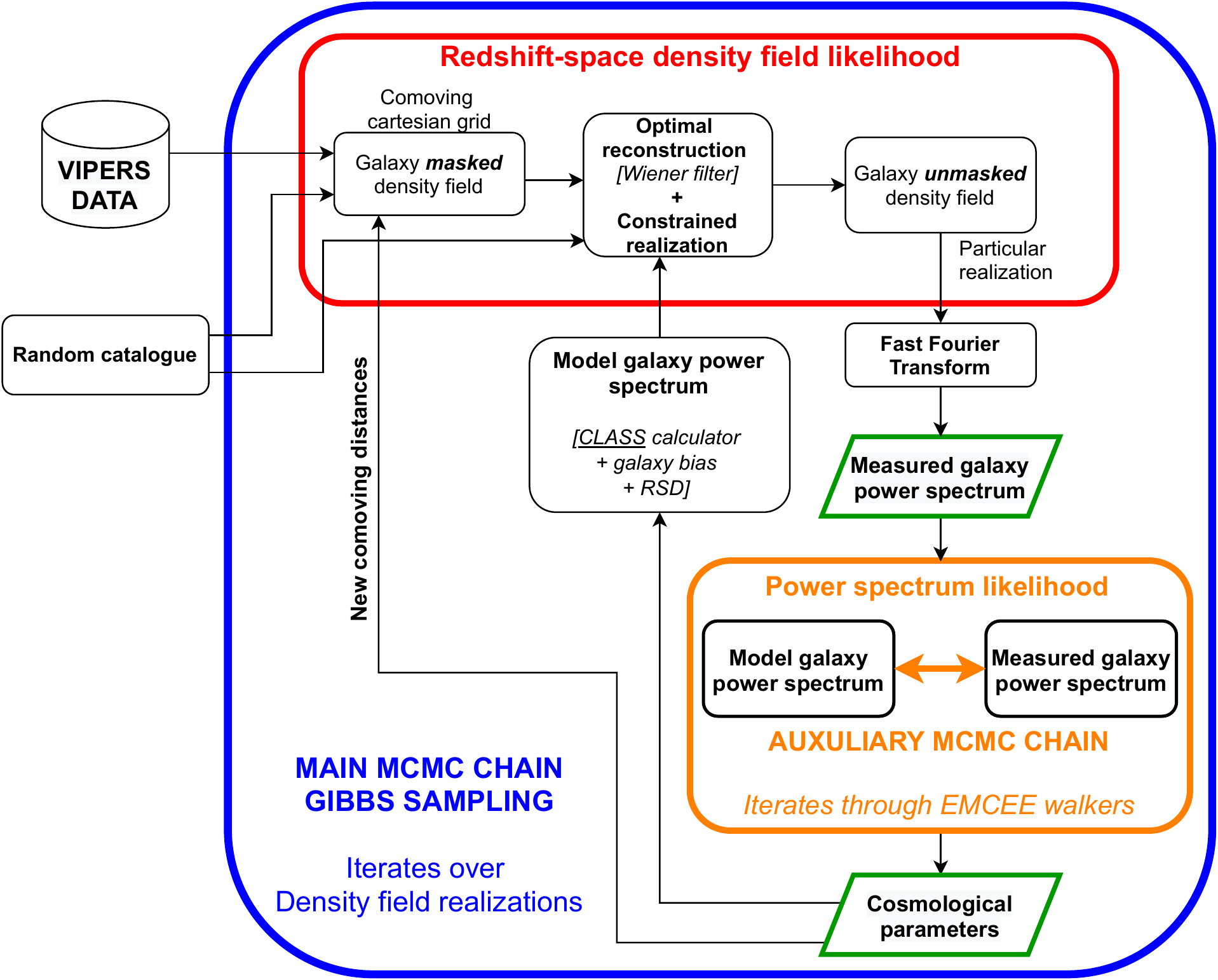}
    \caption{A flow chart summarising the structure of our code. \textcolor{black}{The input includes VIPERS data (RA, DEC, z, TSR and SSR) and the random catalogue, which is generated using the angular mask and redshift distribution, and accounts for the mean density by cell.} The Gibbs sampler is represented by the external blue box. The first section of the chain, described by equation \ref{eq:gibbs1} is represented by the red box (Section \ref{sec:wienerfilter}). The second section of the chain, corresponding to equation \ref{eq:gibbs2} is represented by the orange box (Section \ref{sec:emcee}). The outputs of our code are the galaxy power spectrum and the set of cosmological parameters (internal green boxes).}
    \label{fig:gibbs_diagram}
\end{figure*}
\subsection{Results on VIPERS real and mock catalogues}
\label{sec:resultsonmocks}

We now describe the application of the Gibbs sampling algorithm to the masked data from VIPERS and corresponding mock catalogues. The mock and real catalogues are treated identically, except for the fact that the TSR and SSR completeness corrections are not applied to the mocks. The input data catalogue contains the following galaxy parameters: right ascension, declination, redshift, quality flags (including TSR and SSR weights for real galaxies) and the polygon mask of the survey. Briefly, the steps of the analysis can be summarised as follows.
\begin{itemize}
    \item \textit{Step 1}: Computing the \textit{masked} galaxy density field on the Cartesian comoving grid. \textcolor{black}{Since we adopt a $\Lambda$CDM model, only a} starting value for $\Omega_M$ is needed to compute the comoving distances of the galaxies. In subsequent iterations the value of $\Omega_M$\textcolor{black}{, and so the comoving distance for each galaxy,} is updated for consistency. 
    \item \textit{Step 2}: Generating a particular realization of the \textit{unmasked} galaxy density field with the Wiener filter and constrained realization.
    \item \textit{Step 3}: Computing the power spectrum of the particular realization of the galaxy density field.
    \item \textit{Step 4}: Using the MCMC {\tt Emcee} sampler, generate samples of the model parameters (Eq. \ref{eq:likelihoodemcee}).
\end{itemize}
The output of the Gibbs sampler on each iteration step is the set of model parameters, the measured power spectrum and the realization of the density field. 

The analysis of the VIPERS data was iterated for 1000 steps, while 600 steps were used for the mock catalogs. The first 500 steps (200 for mocks) were discarded as the burn-in period. \textcolor{black}{The main purpose of computing our algorithm on the mocks is to estimate the dispersion of our results, so we can use a lower number of steps, saving computational resources.}

\begin{figure*}
	\includegraphics[width=18cm]{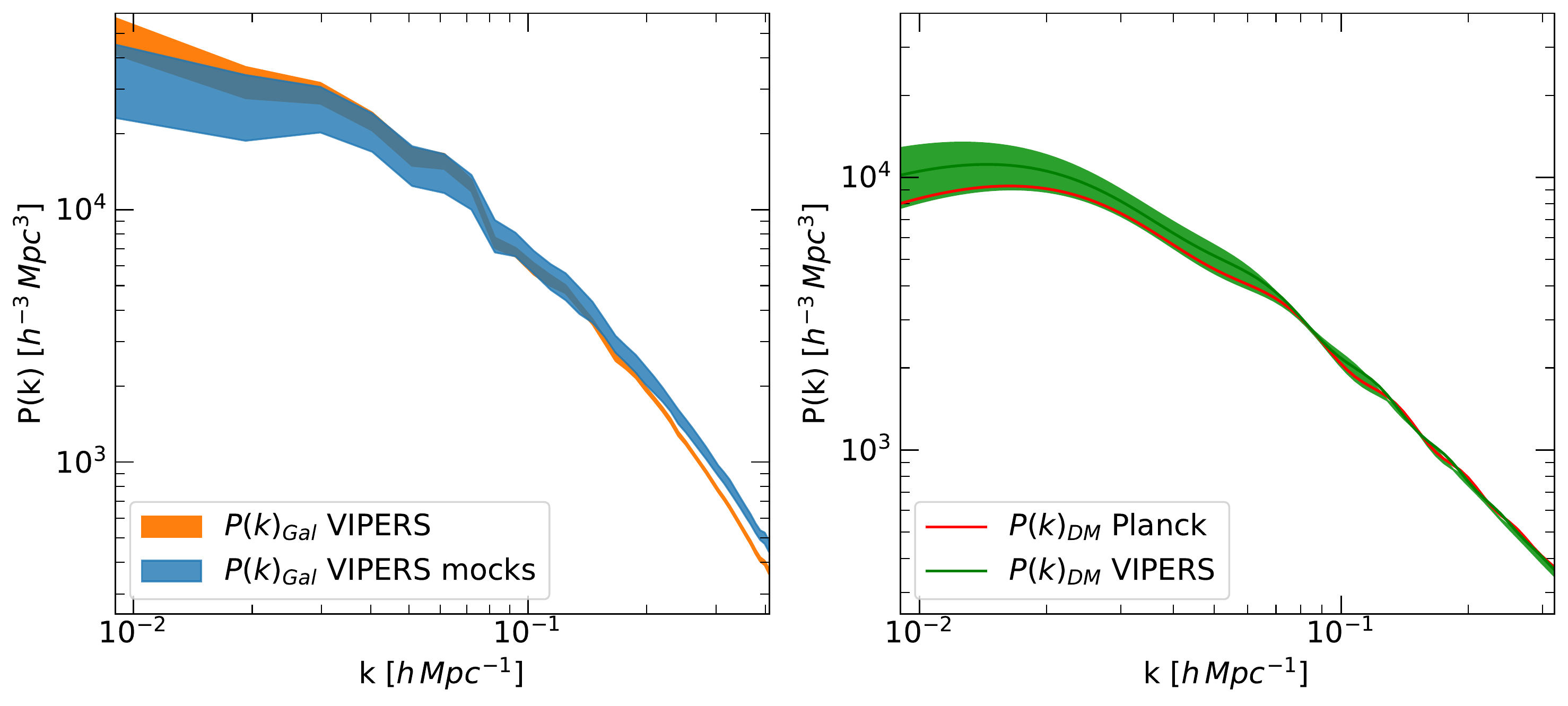}
    \caption{The power spectrum computed for the whole VIPERS sample, W1 and W4 fields using the redshift interval $0.6 < z < 1.0$. The left panel shows  the power spectrum monopole in redshift-space measured from the unmasked galaxy density field. The right panel  shows the constraints on the model  dark matter power spectrum in real space computed with the parameters from the Gibbs sampler.}
    \label{fig:viperspower}
\end{figure*}
\begin{figure*}
\center
	\includegraphics[width=12cm]{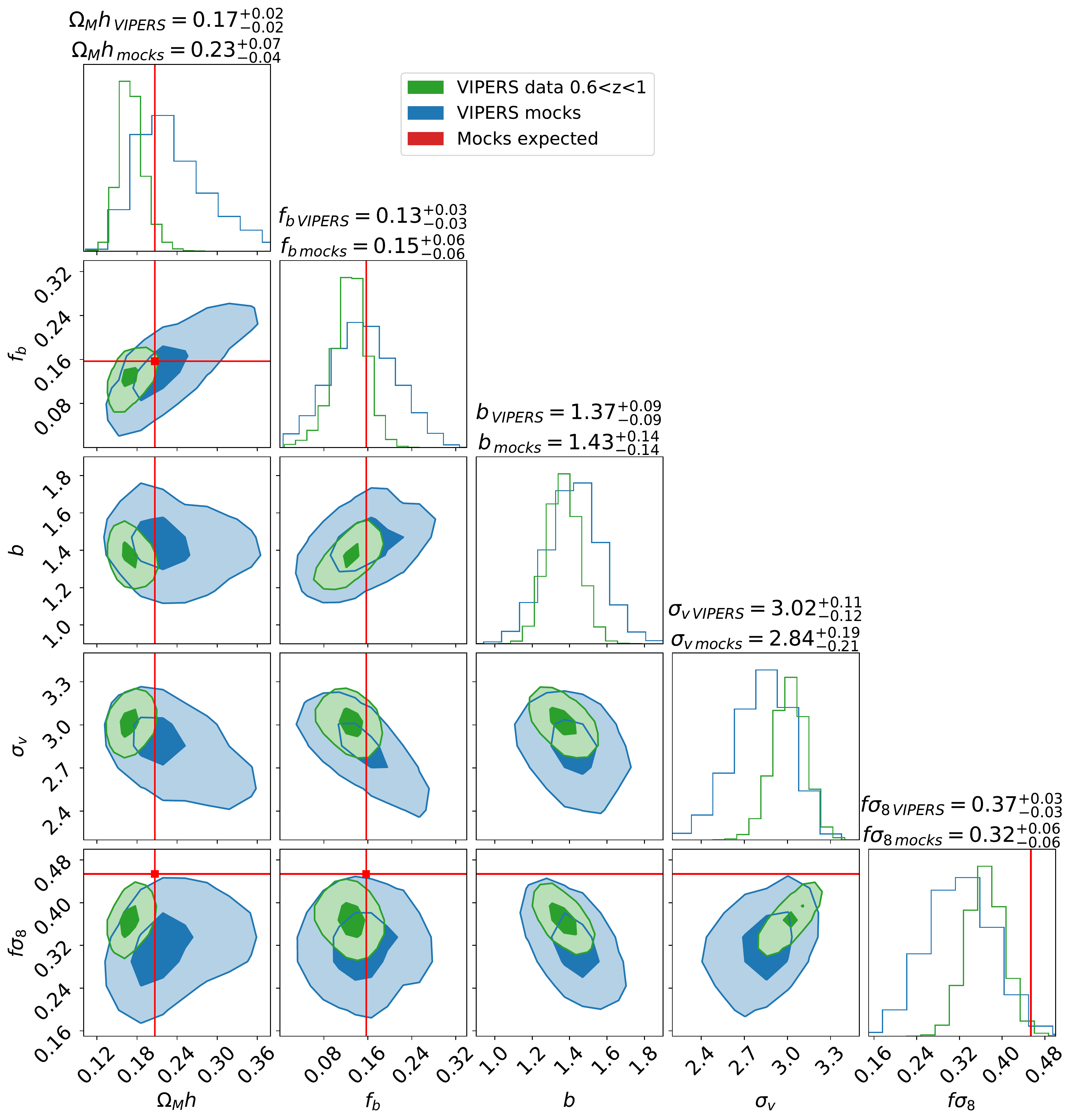}
    \caption{Joint probabilities obtained with the Gibbs sampler applied to the VIPERS data and mock catalogues. The green contours are the results obtained with the VIPERS data. The blue contours are obtained from the mock catalogues: the mean of the mocks distribution provides an estimate of the accuracy of our algorithm, while its dispersion gives a realistic estimate of the precision of our results. In this respect the mean of the mock estimates is determined with a precision that is $\sim\sqrt{153}$ times smaller than the uncertainty shown by the blue contours (which represent a single realisation). The red lines are the values used to generate the mocks according to the BigMD simulation using Planck values. \textcolor{black}{ The contours correspond to the 39th and 86th percentiles.}}
    \label{fig:corner_VIPERS_MOCKS}
\end{figure*}
\begin{figure}
    \includegraphics[width=\columnwidth]{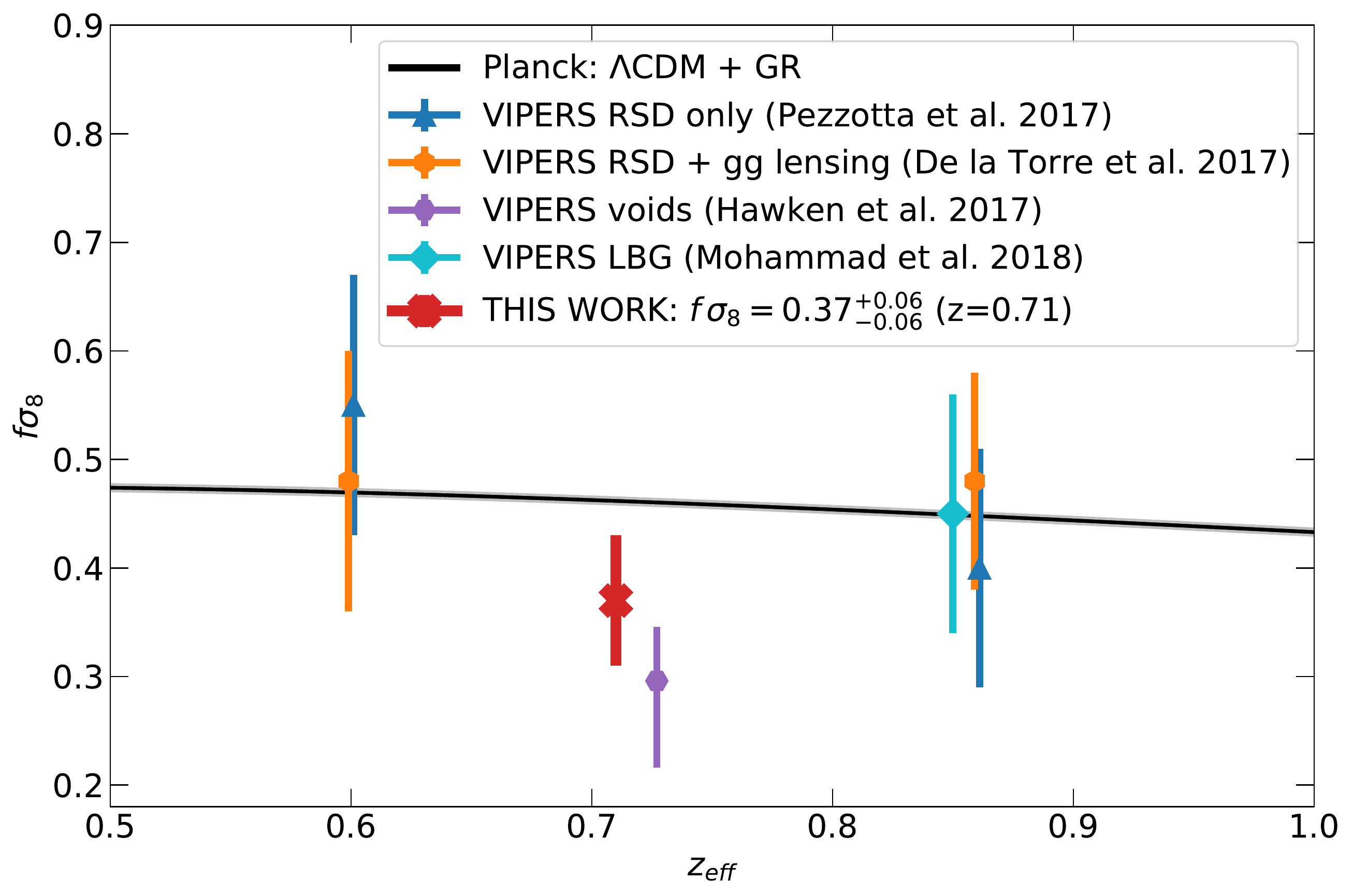}
    \caption{Estimate of the linear growth rate $f\sigma_8$ obtained here from the complete VIPERS galaxy sample (red marker), compared to other VIPERS measurements using different techniques. The black solid curve shows predictions of General Relativity with a $\Lambda$CDM model with parameters set to Plank 2018 \citep{planck18}. }
    \label{fig:fsig8}
\end{figure}

\subsection{Galaxy power spectrum}
The left panel of Fig. \ref{fig:viperspower} shows the galaxy power spectrum monopole in redshift space, measured from the unmasked galaxy density field. The shaded region indicates the 68\% confidence interval estimated from the samples of the Gibbs chain. The results from the mocks and VIPERS data are shown in orange and blue, respectively. Excellent agreement is found between the mocks and VIPERS data at $k<0.2$. At higher $k$ the mocks show a higher power spectrum amplitude than the data. We attribute this divergence to a systematic enhancement in the velocity dispersion in the mocks compared with the data.  This excess of velocities along the line of sight is reflected in the loss of power at small scales. We also note that the confidence interval is significantly smaller in the data compared with the mocks; this is due to the fact that the Gaussian likelihood underestimates the covariance which is instead captured by the scatter in the estimates from the 153 mock realisations. \textcolor{black}{The same trend was observed in \citep{granett15}; however, in that work the galaxy sample was divided into many sparser subsamples each with higher levels of shot noise that led to the overall more Gaussian behavior.} 
 
We next consider the power spectrum model evaluated with the parameters obtained with the Gibbs sampler. The constraints on the parameters themselves will be presented in the following section. The right panel of Fig. \ref{fig:viperspower} shows the posterior region of the non-linear matter power spectrum model in real space. The green shaded area shows the 68\% confidence interval found with the VIPERS data, which shows agreement with the Planck model. The phase shift that can be seen in the Baryon acoustic oscillations is an artefact of the power spectrum model. The BAO signal is not detected in the VIPERS power spectrum, as seen in the left panel, due to the survey geometry \citep[see also the discussion of the window function in][]{rota17}. Although we have successfully \textcolor{black}{evaluated an unmasked density field }in this analysis, the anisotropic shape of the window leaves a fundamental uncertainty in the estimate of the underlying power spectrum that washes out the BAO signal.

\begin{figure}
    \includegraphics[width=\columnwidth]{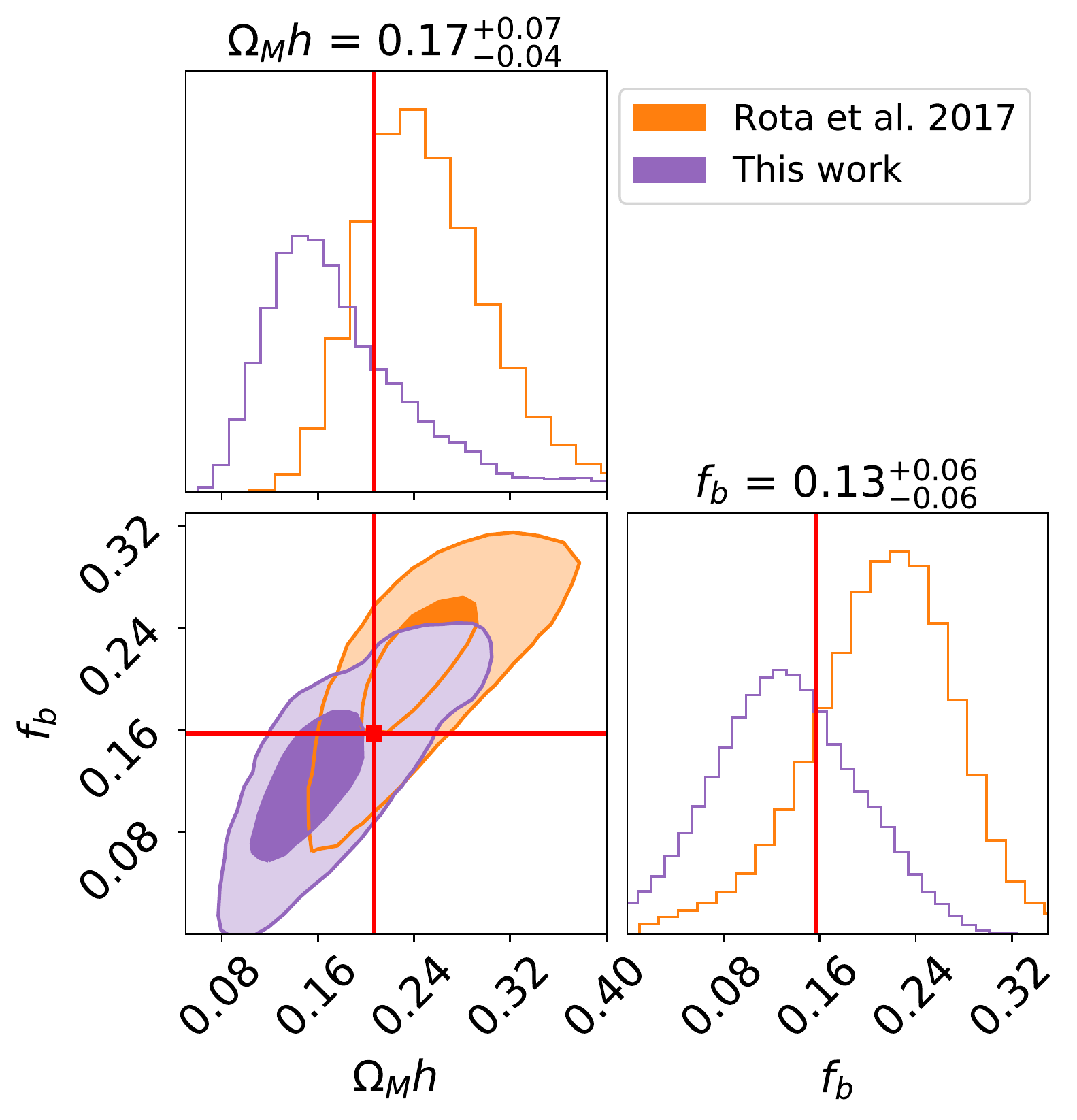}
    \caption{Comparison between the conditional probabilities on $\Omega_Mh$ and $f_b$ between this work and \citet{rota17}. \textcolor{black}{The shape of our contours (magenta) is given by the mocks while the mean value is given by the real VIPERS data.} We find a 1$\sigma$ shift in the contours with respect to Rota which we discuss in Sec. \ref{sec:cosmo}. The Planck best-fit parameters are indicated by the horizontal and vertical lines. \textcolor{black}{ The contours correspond to the 39th and 86th percentiles.}}
    \label{fig:rota_nic}
\end{figure}
\label{sec:resultsonvipers}

\subsection{Cosmological parameters}\label{sec:cosmo}
In Fig. \ref{fig:corner_VIPERS_MOCKS} we show the joint probabilities on cosmological parameters from the Gibbs chains computed on VIPERS (green) and  mock catalogues (blue). The blue contours represent the concatenation of the chains from 153 mock catalogues after excluding the burn-in steps. The red lines are the values used to generate the mocks, in accordance to Planck results. We do not have reference values for the velocity dispersion, $\sigma_v$, and galaxy bias, $b$, because these parameters are determined by the N-body simulation, the HOD model and the galaxy sample selection and cannot be precisely predicted analytically.

We find that the analysis on the mocks recovers the values of the matter density, $\Omega_Mh$, and Baryon fraction, $f_b$, demonstrating the soundness of the analysis. However, the growth rate of structure is underestimated at a level of 2.2$\sigma$. 

Turning to the VIPERS data, we constrain the matter density to $\Omega_Mh= 0.17^{+0.07}_{-0.04}$. The error range given here was inferred from the scatter in the estimates from the mocks. The choice of sampling $\Omega_Mh$ provides us with the opportunity of comparing directly the results with previous VIPERS measurements, in particular those obtained by \cite{rota17} through the traditional FKP estimator: $\Omega_Mh= 0.227^{+0.063}_{-0.050}$. This small tension at the 1$\sigma$ level, visible in the figure \ref{fig:rota_nic}, will be discussed below.

The Baryon fraction obtained in this work is $f_B= 0.13^{+0.06}_{-0.06}$. Again we find an agreement with respect to \cite{rota17}: $f_B= 0.220^{+0.058}_{-0.072}$. A comparison of the joint likelihood of the matter density and baryon fraction is plotted in Fig. \ref{fig:rota_nic}. The size of the contour that we find is similar to \cite{rota17} but shifted by 1$\sigma$.

The galaxy bias is  $b= 1.37^{+0.14}_{-0.14}$. The 1$\sigma$ interval for this value overlaps with the $b_{eff}$ found by \cite{granett15}: $b_{eff}=1.44^{+0.02}_{-0.02}$ using the same redshift interval and agrees with previous VIPERS analyses \citep{2013A&A...557A..17M,2016A&A...594A..62D}.
The velocity dispersion along the line of sight is mesured to be $\sigma_v= 3.02^{+0.19}_{-0.21}$ which matches the value found with the mocks, when accounting for redshift measurement errors as seen in Fig. \ref{fig:corner_VIPERS_MOCKS}.

The anisotropic galaxy power spectrum can uniquely constrain the parameter combination $f(z)\sigma_8(z)$ in linear theory. We find $f\sigma_8 = 0.37^{+0.06}_{-0.06}$ at \textcolor{black}{$z=0.71$}, as shown in Fig. \ref{fig:fsig8} together with the theory predictions in Planck cosmology. Our measurement is lower than predicted and also lower than previous VIPERS measurements based on two-point correlation analysis \citep[i.e.][]{pezzotta17,Delatorre17,mohammad18}. This could be due to having modelled RSD using the dispersion model, which is known to underestimate the growth rate \citep{mohammad18}. \textcolor{black}{The VIPERS reference RSD measurements were made on the correlation function, instead here we fit the power spectrum model.  We applied a maximum scale $k_{max}=0.4 h {\rm Mpc}^{-1}$ which corresponds to a spatial scale of $\pi/0.4\sim0.8 h^{-1}\rm{Mpc}$, although it is not equivalent to the minimum scale in the correlation which prevents us from making a direct comparison.} We note however, that our estimate has a precision of \textcolor{black}{$\Delta f\sigma_8/f\sigma_8=0.16$, in comparison to the previous VIPERS measurements from  \citet{pezzotta17} in the redshift bin $0.5<z<0.7$, $\Delta f\sigma_8/f\sigma_8=0.22$ and $0.7<z<1.2$, $\Delta f\sigma_8/f\sigma_8=0.27$. Our analysis gives better precision despite having greater freedom in the model due to the fact that we have jointly estimated the matter density parameter and it is not subject to geometric distortions \citep{2012MNRAS.426.2566M, Ballinger96}.}

\textcolor{black}{To test the robustness of the constraints we performed two additional analyses on  VIPERS with narrow priors imposed on selected parameters.  First, to investigate whether the low value of $f\sigma_8$ is related to the freedom of the modelling, we placed narrow priors on $\Omega_m$ and $f_b$, leaving the bias and RSD paramters $f\sigma_8$ and $\sigma_v$ free. The results of this more constrained configuration are equal to the original findings with $f\sigma_8=0.37^{+0.03}_{-0.03}$ and have consistent values of bias and $\sigma_v$. We conclude from this exercise that the low value of $f\sigma_8$ is robust to the modelling of the shape of the power spectrum.}

\textcolor{black}{As a second robustness test, we imposed the prior $0.43<f\sigma_8<0.46$ and left the other parameters free. This allows us to investigate the 1$\sigma$ tension with respect to \citet{rota17} in the estimates of $\Omega_m h$ and $f_b$ seen in Fig. \ref{fig:rota_nic}.} This configuration better matches the analysis of Rota which only used the power spectrum monopole.

\textcolor{black}{In this configuration we found a higher value of the matter density $\Omega_m h=0.19$ to be compared with $\Omega_Mh= 0.227^{+0.063}_{-0.050}$ from \citet{rota17}; thus, the narrow prior reduces but does not solve the tension.}
This shows that the results of the Gibbs sampler are stable within 1$\sigma$; however, the joint fit of redshift-space distortions and the shape of the power spectrum prefers lower values of both $\Omega_mh$ and $f\sigma_8$.

\section{Discussion and conclusions}
\label{sec:conclusions}
We present a complete analysis of the final data release of VIPERS, based on a maximum likelihood estimator for the density field and cosmological parameters. Our primary aim was  to improve the accuracy of previous VIPERS results through the use of an estimator that optimally corrects for the complex survey geometry. We reconstructed the un-masked redshift-space galaxy density field using the Wiener filter, which is the maximum posterior estimator in the case of a Gaussian field. With the Gibbs sampling approach we estimated the joint posterior probabilities on the parameters of the $\Lambda$CDM model as shown in Fig. \ref{fig:corner_VIPERS_MOCKS}. We summarise our results as follows:
\begin{enumerate}
    \item \textcolor{black}{Starting from a Bayesian formalism described in \citet{granett15} }we developed \textcolor{black}{a new} analysis pipeline for spectroscopic galaxy surveys to estimate the joint posterior probabilities of the density field in redshift-space along with the cosmological parameters.
    \item We applied the pipeline to 153 mock realisations of VIPERS and found the \textcolor{black}{matter and baryon density} parameter estimates to be unbiased. \textcolor{black}{In both the mocks and data, the estimates of the growth-of-structure parameter $f\sigma_8$ is systematically low.  We attribute this systematic bias to the use of the dispersion model for the anisotropic power spectrum.}
    \item When applied to the VIPERS data, the algorithm yields the un-masked density field in redshift-space and its power spectrum monopole.
    \item We derived parameter constraints on the matter density, baryon fraction and growth of structure parameter estimating their uncertainity from the dispersion in the mocks.
    \item The precision we find on the cosmological parameters is competitive with previous VIPERS analyses \citep[\textcolor{black}{the former for $\Omega_M$ and $f_b$, the latter for $f\sigma_8$}]{rota17,pezzotta17}, despite the larger number of degrees of freedom, which here include the shape of the power spectrum \textcolor{black}{and} redshift-space distortions. \textcolor{black}{The analysis does not sufffer from Alcock-Pacynzki geometric distortions since the density field is recomputed consistently on every iteration.}
    \item Our constraint of the growth-of-structure parameter $f\sigma_8$ is systematically low, which we attribute to the use of the dispersion model for the anisotropic power spectrum.  \textcolor{black}{We also found $f\sigma_8$ and $\Omega_Mh$ to be correlated and the analysis of the VIPERS data also preferred a low value of $\Omega_Mh$ with a 1-$\sigma$ tension with \citet{rota17}.  Fixing $f\sigma_8$ with a tight prior decreased this tension. 
    However, given the volume of the VIPERS data set, these results are consistent with the statistical error, so a larger data set is needed to investigate the issue further.}
    \item Our results on the matter density and baryon fraction parameters are in agreement with measurements at lower redshift from 2dFGRS at $z=0.2$ \citet{cole05}, SDSS LRG at $z=0.35$ \citet{tegmark04}, and WiggleZ at $0.2<z<0.8$ \citet{parkinson12} as well as with Planck determinations \citep{planck18}.
\end{enumerate}

The maximum-likelihood analysis we have presented represents a forward model of the galaxy survey based on a multivariate Gaussian likelihood and prior for the density field (Eq. \ref{eq:likelihoodemcee}, \ref{eq:prior}). This model could be modified to account for the notably non-Gaussian distribution of the density field, for example with a log-normal distribution  \citep{Kitaura2010} and non-Poissonian sampling models \citep{Ata2015}. It has been shown that more cosmological information  can be unlocked from the field with the log-normal model than is available in the standard two-point statistics of the density field \citep{Carron2014}. However, limiting ourselves to the 
Gaussian case, we found that our error estimates were underestimated with respect to the dispersion of the mock catalogues. This aspect of the algorithm may be improved by adopting a prior and likelihood that better represent the galaxy distribution on the quasi-linear scales that we consider. Even so, the discrepancy in the error analysis is likely to be less significant for larger surveys for which coarser grids can be employed for the density field reconstruction.

To optimise the use of upcoming surveys it will be necessary to carry out joint analyses between multiple observables and account for a multitude of observational systematic effects. Forward-modelling approaches at the level of the density field provide a promising approach. However, more experience is needed to build and apply forward modelling to the next generation of galaxy surveys. The application to VIPERS is one step in this direction.

\section*{Acknowledgements}
N.E. acknowledges F. Tosone, E. Sarpa, E. Branchini, C. Carbone and G. Verza for helpful discussions during the preparation of this work. This paper uses data from the VIMOS Public Extragalactic Redshift Survey (VIPERS). VIPERS has been performed using the ESO Very Large Telescope, under the "Large Programme" 182.A-0886. The participating institutions and funding agencies are listed at \url{http://vipers.inaf.it}. This work has been financially supported by Italian MUR PRIN 2017, grant n.20179P3PKJ. 

\section*{Data Availability}
The data underlying this article are available in \url{http://vipers.inaf.it}. The code developed for this work will be shared on reasonable request to the corresponding author.

\bibliographystyle{mnras}
\bibliography{bib_masterthesis}


\bsp	
\label{lastpage}
\end{document}